\documentclass[prd, aps, nofootinbib, preprintnumbers, showpacs, superscriptaddress, twocolumn]{revtex4}

\usepackage{graphicx}
\usepackage{bm}
\usepackage{amsmath}
\usepackage{amsfonts}

\def\de{\partial} 

\begin{document}

\title{Faithful Effective-One-Body waveforms of \\
small-mass-ratio coalescing black-hole binaries}

\author{Thibault Damour}
\affiliation{Institut des Hautes Etudes Scientifiques, 35 route de Chartres, 91440 Bures-sur-Yvette, France}

\author{Alessandro Nagar}
\affiliation{Dipartimento di Fisica, Politecnico di Torino, Corso Duca degli Abruzzi 24,
              10129 Torino, Italy and INFN, sez. di Torino, Via P.~Giuria 1,
              Torino, Italy}

\begin{abstract}
We address the problem of constructing high-accuracy, faithful analytic  waveforms
describing the gravitational wave signal emitted by inspiralling and coalescing
binary black holes. We work within the Effective-One-Body (EOB) framework and
propose a methodology for improving the current (waveform) implementations of
this framework based on understanding, element by element, the physics behind 
each feature of the waveform, and on systematically comparing various
EOB-based waveforms with ``exact'' waveforms obtained by numerical relativity 
approaches. The present paper focuses on  small-mass-ratio non-spinning binary
systems, which can be conveniently studied by Regge-Wheeler-Zerilli-type
methods. Our results include: (i) a resummed, 3PN-accurate description of the 
inspiral waveform, (ii) a better description of radiation reaction during the 
plunge, (iii) a refined analytic expression for the plunge waveform, (iv) an 
improved treatment of the matching between the plunge and ring-down
waveforms. This improved implementation of the EOB approach allows us to
construct complete analytic waveforms which exhibit a remarkable 
agreement with the ``exact'' ones in modulus, frequency and phase.
In particular, the analytic and numerical waveforms stay in phase, during the
whole process, within $\pm 1.1 \%$ of a cycle. We expect that the extension
of our methodology to the comparable-mass case will be able to generate 
comparably accurate analytic waveforms of direct use for the ground-based
network of interferometric detectors of gravitational waves.

\end{abstract}

\date{\today}


\maketitle

\section{Introduction}
\label{sec:intro}
Coalescing black hole binaries are  among the most promising
gravitational wave (GW) sources for the currently operating ground-based 
detectors like GEO/LIGO/VIRGO. The most useful part of the
waveform for detection comes from the most relativistic part of the dynamics:
the last few cycles of the adiabatic
inspiral, the plunge and the merger.
Since LIGO is currently taking data at the expected sensitivity, it becomes
urgent to have accurate template waveforms for detection. In the most general,
spinning case,
these waveforms are complicated functions of the initial masses $m_1$ and 
$m_2$ and the spins ${\bf S}_1$ and ${\bf S}_2$ of the two constituent black
holes. Due to the multi-dimensionality of the parameter space, it seems impossible
for state-of-the-art numerical simulations to densely sample this parameter
space. This motivates a renewed effort for developing {\it analytical} methods 
towards computing the 
 huge banks of accurate template waveforms needed to densely cover the 
parameter space.

As far as we know, the first  estimate of the complete waveform (covering
inspiral, plunge, merger and ring-down)
of a coalescing,  non-spinning black hole binary was made  
 in 2000~\cite{Buonanno:2000ef}, on the basis of a new 
analytical approach to the general relativistic two-body dynamics, the 
{\it Effective-One-Body} (EOB) approach~\cite{Buonanno:1998gg}. At the time,
 no reliable numerical simulations of coalescing black hole binaries
were yet available. This EOB-based estimate used a 2.5 post-Newtonian (PN)
accurate description of the dynamics (the only one available at the 
time\footnote{This EOB waveform estimate was later updated by taking into
account the more accurate 3.5 PN dynamics, as well as by considering
spinning black holes~\cite{Buonanno:2005xu}.})
 and a quadrupole-type ``restricted waveform''
approximation for the waveform down to a ``matching radius'' $r_{\rm match}$
where the two-black-hole system was replaced by a unique, ringing Kerr
black hole. It was also argued that the matching radius $r_{\rm match}$ could be taken as the
EOB-deformed ``light ring'', $r_{\rm LR}$, which was noticed to be very close to the point
where the orbital frequency reached a maximum.
 As a first approximation, a complete waveform was computed by
matching, at $r_{\rm match} \simeq r_{\rm LR}$, the inspiral-plus-plunge   waveform 
to a ring-down waveform made of
 the least damped quasi-normal mode (QNM) of a Kerr black hole of mass 
and angular momentum corresponding to those of the relative dynamics at
$r_{\rm match}$. In other words, Ref.~\cite{Buonanno:2000ef} was making the
assumption that the {\it merger phase} of the two black holes was very brief,
and did not correspond to any especially strong GW emission feature, but would
simply smoothly connect the plunge behavior to the ringdown one.

In 2001, a combination of (rather short) full numerical simulations
with a ``close-limit'' approximation~\cite{Price:1994pm} led to
 the first numerical estimate of the waveform emitted by the plunge
 from the ISCO, followed by merger and ringdown~\cite{Baker:2001nu,Baker:2002qf}.
Very recent breakthroughs in numerical relativity simulations have finally
made accessible a much more precise knowledge of
the waveforms generated by the merger of two black holes
of comparable masses $m_1$ and $m_2$, possibly with spin.
These results have been obtained by different groups with different approaches,
and exhibit convincing internal convergence and a nice mutual 
consistency~\cite{Pretorius:2005gq,Campanelli:2005dd,Diener:2005mg,Campanelli:2006gf,
Baker:2006yw,Baker:2007fb,Campanelli:2006uy,Gonzalez:2006md,Thornburg:2007hu,Koppitz:2007ev}.
The availability of reliable numerical results makes it urgent to compare
analytical and numerical approaches for (i) understanding in depth the
physics that is involved in the process and thus (ii)  completing the 
currently available analytical knowledge of the problem  with the ``missing'' 
non-perturbative physics provided by numerical
relativity simulations.  Several different ways of combining the  knowledge
acquired through analytical and numerical methods have been recently explored,
such as the use of ``hybrid'' signals made by stitching together an inspiral PN-type waveform 
to a NR merger one~\cite{Pan:2007nw,Ajith:2007qp}, or the study of whether some simple 
multiparameter (BCV-type~\cite{Buonanno:2002ft}) analytical waveform can have 
sufficiently large overlaps with hybrid signals. Our aim here is different.
We wish, ultimately, to construct 
{\it high-accuracy} purely analytical template waveforms, depending only
on physical parameters, and covering the full process: 
inspiral, plunge, merger and ringdown; by high-accuracy we mean that their 
phasing (and possibly their amplitude too) would be very close to the true phasing, 
uniformly, to a few percent. Let us recall that it is convenient
(following the terminology of Ref.~\cite{Damour:1997ub}) to distinguish
 {\it effectual} and {\it faithful} templates: effectual templates are
 defined by the property of having sufficiently large (say $\geq 96.5 \%$)
 overlaps with the
 expected ``real'' signals, {\it after maximization over all (kinematical,
 dynamical, and possibly fudge) parameters}. Effectual templates maybe useful tools
 for {\it detecting} real signals, but they might lead to large biases in
 the measurement of the physical parameters of the system. By contrast,
faithful templates are defined by the property of having not only large
overlaps, but also of being so ``close'' to the real signals that the
overlap between the template and the real signal is maximized for values
of the parameters which are very close to the real ones (``small biases'').
Recent ``first order'' comparisons between full numerical relativity waveforms
and EOB-based waveforms have already established that EOB waveforms can be
{\it effectual} (overlaps $\geq 96.5 \%$) over the full mass
range of interest for GEO/LIGO/VIRGO, say a total mass
$M = m_1 + m_2$ between $10$ and $120$ solar masses~\cite{Buonanno:2006ui,Pan:2007nw}.
Our main aim will be to prove that EOB-based waveforms can also be {\it faithful}
over the full mass range of interest.
Our methodology for achieving this aim is to improve the current waveform 
implementations~\cite{Buonanno:2000ef,Buonanno:2005xu,Buonanno:2006ui,Pan:2007nw}
of the EOB approach~\cite{Buonanno:1998gg,Buonanno:2000ef, Damour:2000we,Damour:2001tu,Damour:2006tr}
by understanding, element by element, the physics behind each feature of 
the waveform, and by comparing EOB-based waveforms with ``exact'' waveforms 
obtained by numerical relativity approaches. In the
present paper, we shall tap the information contained in small-mass-ratio 
waveforms. [A preliminary version of our results was reported 
in~\cite{Damour:2006qz}.] Indeed, the limiting case $m_1m_2\ll (m_1+m_2)^2$ 
constitutes a ``clean laboratory'' where many subtle physical issues can 
be studied in detail (and by means of rather light numerical tools), without the
complications entailed by the current three-dimensional numerical
simulations (residual eccentricity, numerical noise,...). We rely on
the recently developed numerical waveform computations \`a la Regge-Wheeler-Zerilli in 
Ref.~\cite{Nagar:2006xv}. Note that we consider non-spinning inspiralling systems
circularized by radiation reaction.

The various improvements in the EOB waveform construction that we shall address
here concern: (i) an improved analytical expression for the ($(\ell,m) =(2,2)$)
waveform which includes a resummation of the tail effects, and a 3PN-accurate 
``non-linear'' amplitude correction, (ii) the inclusion of non-quasi-circular corrections
to the waveform, (iii) a comparative study of the two radiation reaction
expressions during the plunge that have been proposed (\cite{Buonanno:2000ef}
versus \cite{Damour:2006tr}), (iv) the inclusion of non-quasi-circular corrections
to radiation reaction,  and (v) an improved treatment of the matching between the plunge
and ring-down waveforms (which takes into account a new understanding of the
importance of the number of QNMs, the sign of their frequencies, and the 
length of the interval on which the matching is done). We shall show that the resulting
improved implementation of the EOB approach yields very faithful
waveforms whose amplitude and phase agree remarkably well with the ``exact'' ones:
in particular, the EOB phasing differs from the ``exact'' one by less than
$\pm 1.1 \%$ of a cycle over the whole process.

This paper is organized as follows. In Sec.~\ref{sec:dynamics} we recall
the EOB construction for the relative dynamics of the binary system 
and specify it to the small mass ratio limit. In Sec.~\ref{sec:waveforms}
we discuss the computation of the GW emission and we focus on the
comparison between different expressions of radiation reaction. 
Sec.~\ref{sec:eob_waves} introduces our ``best'' EOB-type waveform 
for the transition from inspiral to plunge and shows that it is an excellent 
approximation to the ``actual'' waveform. Some conclusions are presented 
in Sec.~\ref{sec:conc}.

\section{EOB relative dynamics}
\label{sec:dynamics}
In this section we recall the EOB non-perturbative construction of the two-body 
dynamics including a radiation reaction force and then specify it to the
small mass limit.
The EOB approach to the general relativistic two-body dynamics is a 
{\it non-perturbatively resummed}  analytic technique which has been developed in 
Refs.~\cite{Buonanno:1998gg,Buonanno:2000ef,Damour:2000we,Damour:2001tu,Buonanno:2005xu,Damour:2006tr}.
This technique uses, as basic input,  the results of PN theory,
such as: (i) PN-expanded equations of motion for two point-like bodies,
 (ii) PN-expanded radiative multipole moments, and (iii) PN-expanded
energy and angular momentum fluxes at infinity. For the moment, the
most accurate such results are the 3PN conservative 
dynamics~\cite{Damour:2001bu,Blanchet:2003gy}, and the 3.5PN energy 
flux~\cite{Blanchet:2001aw,Blanchet:2004bb,Blanchet:2004ek}. Then
the EOB approach ``packages'' this PN-expanded information in special 
{\it resummed} forms which extend the validity of the PN results beyond 
the expected weak-field-slow-velocity regime into (part of) the 
strong-field-fast-motion regime.
At the practical level, the result of this ``packaging'' is that the 
complicated PN-expanded relative dynamics, (in the center of mass frame) 
of a binary system of masses $m_1$ and $m_2$ is mapped into the simpler 
(essentially geodesic) dynamics of a particle of mass $\mu=m_1m_2/(m_1+m_2)$ moving 
in some effective background geometry (in Schwarzschild gauge)
\begin{equation}
ds^2 = -A(r)dt^2+B(r)dr^2 + r^2(d\theta^2+\sin^2\theta d\varphi^2) \ .
\end{equation}
Here and below we work with dimensionless reduced variables $r=R/M$ and
$t=T/M$, with $M=m_1+m_2$; $(r,\theta,\varphi)$ are polar coordinates
in the {\it effective} problem that describe the relative motion.
The dynamics of the effective particle is described by a Hamiltonian 
$H_{\rm EOB}(M,\mu)$ and a radiation reaction force ${\cal F}_{\rm EOB}(M,\mu)$.
In the general comparable-mass case $H_{\rm EOB}$
has the structure $H_{\rm EOB}(M,\mu)=M\sqrt{1+2\nu(\hat{H}_\nu - 1)}$
where $\nu\equiv \mu/M\equiv m_1m_2/(m_1+m_2)^2$ is the symmetric mass ratio.
In the test mass limit that we are considering, $\nu\ll 1$, we can expand
$H_{\rm EOB}$ in powers of $\nu$. After subtracting inessential constants
we get a Hamiltonian per unit ($\mu$) mass 
$\hat{H}=\lim_{\nu \to 0}(H_{\rm EOB}-{\rm const.})/\mu=\lim_{\nu\to 0}\hat{H}_\nu$.
It is convenient to replace the Schwarzschild radial coordinate $R$ by the
Regge-Wheeler tortoise coordinate 
$R_*=R+2M\log[R/(2M)-1]$, and, correspondingly, the radial momentum $P_R$ by 
the conjugate momentum $P_{R_*}$ of $R_*$. Then the 
(specific) Hamiltonian reads~\cite{Nagar:2006xv}
\begin{equation}
\label{schw:ham}
\hat{H} = \sqrt{ A(r)\left(1+\frac{p_{\varphi}^2}{r^2} \right)+p_{r_*}^2 } \ .
\end{equation}
Here we have introduced dimensionless variables $r\equiv R/M$, $r_*\equiv R_*/M$,
$p_{r_*} =P_{R_*}/\mu$ 
and $p_\varphi =P_\varphi/(\mu M)$. In the $\nu\to 0$ limit, we have that
 $A(r)=B(r)^{-1}=1-2/r$. We have used the tortoise canonical pair $(r_*,p_{r_*})$ instead of the
``traditional'' one $(r,p_r)$ in the Hamiltonian for two related reasons. On the one
hand, $p_{r_*}$ has a finite limit when $r$ tends to the zero of $A(r)$ 
($r=2$, the event horizon), while $p_r$ diverges there\footnote{This divergence leads,
in particular, to numerical problems in the source of the Regge-Wheeler-Zerilli equation:
some ``exact'' analytical cancellations that are explicit when expressing the source
in terms of $p_{r*}$ (which stays finite), become implicit when using  $p_r$
and numerically tend to yield a problematically divergent source.}. On the other hand, since one of
our goal is to compute the gravitational wave signal from the relative
dynamics, it is very useful to have the (radial-delta-function) source moving quasi-uniformly
towards $ - \infty$ on the doubly-infinite $r_*$ axis, rather than exponentially slowing down and
getting stuck at $r=2M$ (this allows us to cut-off more smoothly the radial motion after some time).
For details of our numerical implementation see~\cite{Nagar:2006xv}.

Hamilton's canonical equations for $(r,r_*,p_{r_*},p_{\varphi})$ 
in the equatorial plane ($\theta=\pi/2$) yield
\begin{align}
\label{eob:1}
\dfrac{dr_*}{dt} & = \dfrac{p_{r_*}}{\hat{H}} \ , \\
\label{eob:2}
\dfrac{dr}{dt}  & = \dfrac{A}{\hat{H}}p_{r_*}                \equiv v_r \ , \\
\label{eob:3}
\dfrac{d\varphi}{dt} & = \dfrac{A}{\hat{H}}\dfrac{p_\varphi}{r^2} \equiv \Omega \ ,\\
\label{eob:4}
\dfrac{dp_{r_*}}{dt} &= -\dfrac{r-2}{r^3\hat{H}}\left[p_{\varphi}^2\left(\dfrac{3}{r^2}-\dfrac{1}{r}\right)+1\right] \ , \\
\label{eob:5}
\dfrac{dp_{\varphi}}{dt}&= \hat{\cal F}_\varphi \ .
\end{align}
Note that the quantity $\Omega$ is dimensionless and represents the orbital frequency
in units of $1/M$.
In these equations the extra term $\hat{\cal F}_{\varphi}$ (of order $O(\nu)$) 
represents the non conservative part of the dynamics, namely the radiation
reaction force. During the {\it quasi-circular  inspiral}, 
a rather accurate expression for $\hat{\cal F}_{\varphi}$ is the 
following Pad\'e-resummed form
\begin{equation}
\label{FK}
\hat{\cal F}^K_{\varphi}\equiv \dfrac{ {\cal F}^K_{\varphi}}{\mu}
=-\dfrac{32}{5}\nu\Omega^{7/3}\dfrac{\hat{f}_{\rm
    DIS}(v_{\Omega}) } {1-\sqrt{3}v_{\Omega}} \ ,
\end{equation}
which is expressed in terms of the PN ordering parameter 
$v_{\Omega}\equiv \Omega^{1/3}$. In this expression, the 
function $\hat{f}_{\rm DIS}$ denotes the ``factored flux function'' 
of~\cite{Damour:1997ub} (scaled to the Newtonian (quadrupole) flux), taken here 
in the $\nu\to 0$ limit. Ref.~\cite{Buonanno:2000ef} (in the comparable
mass case) assumed that the analytical continuation of the 
expression (\ref{FK}) might still be a sufficiently accurate description
of radiation reaction effects during the {\it plunge}.
 On the other hand, the authors of Ref.~\cite{Damour:2006tr} pointed out
that the expression~(\ref{FK})
assumed the continued validity of the usual Kepler law $\Omega^2r^3 \simeq 1$ 
during the plunge. [This is why we label the expression (\ref{FK}) with a 
superscript $K$, for Kepler.] They, however, emphasized that the Kepler 
combination $K=\Omega^2 r^3$ significantly deviates from one after
the crossing of the Last Stable (circular) Orbit (LSO), 
to become of  order of 0.5 at the light ring. 
Ref.~\cite{Damour:2006tr} went on to argue for a different 
expression for the radiation reaction, say $\hat{\cal F}_{\varphi}$ (without 
any superscript) that does not assume Kepler's law. In the $\nu\to 0$ 
limit this new expression reads
\begin{equation}
\label{eq:DGloss}
\hat{{\cal F}}_{\varphi}  \equiv \dfrac{\cal F_{\varphi}}{\mu}= -\dfrac{32}{5}\nu
\Omega^5 r^4 \dfrac{\hat{f}_{\rm DIS}(v_{\varphi})}{1-\sqrt{3}v_{\varphi}} \ .
\end{equation}
where the ordering PN parameter is given by the azimuthal velocity 
$v_\varphi=r\Omega $. Note that the essential difference between
the two possible expressions for the radiation reaction is 
that $\hat{\cal F}_{\varphi}^K\propto \Omega^{7/3}$, while 
$\hat{\cal  F}_{\varphi}\propto \Omega^5 r^4$. For simplicity, in both
possible expressions for the radiation reaction our current estimate
of $\hat{f}_{\rm DIS}$ is obtained by Pad\'e approximating the 2.5 PN results,
a procedure that has already been shown to be quite robust versus
``exact'' numerical results in~\cite{Damour:1997ub}.

Let us finally comment on the numerical values of the parameter $\nu$ that we shall consider.
First, we should consider values such that the $\nu-$dependent corrections in the two-body effective
Hamiltonian $\hat{H}_\nu$, i.e. the $\nu-$dependent contributions in the EOB metric functions
such as $A(r) = 1 - 2 u + 2 \nu u^3 + \cdots$ (where $u \equiv 1/r$), are indeed numerically
negligible. However, we do not wish to consider too small values of $\nu$. Indeed, the
present study of the inspiral, plunge and merger of a mass $\mu$ into a Schwarzschild black hole
of mass $M$ is intended as a toy model of the coalescence of two comparable-mass black holes.
We therefore wish to minimize the quantitative difference between these two types of inspiralling
motions. In particular, it has been shown in Ref.~\cite{Buonanno:2000ef} that the number
of orbits during the comparable-mass plunge was roughly $\sim ( 4 \nu)^{-1/5}$, i.e. a number
of order one when $m_1=m_2$ (i.e. $\nu = 1/4$). We should {\it not} therefore consider
values of $\nu$ so small as to make $( 4 \nu)^{-1/5}$ much larger than one.
 As a compromise we shall always consider here the value $\nu = 0.01$, which gives only
 $\sim ( 4 \nu)^{-1/5} \simeq 1.9$ orbits during the plunge (as will indeed appear in 
 Fig.~\ref{fig:fig1} below).

\section{Regge-Wheeler-Zerilli waveforms}
\label{sec:waveforms}
\begin{figure}[t]
  \begin{center}
   \includegraphics[width=85 mm, height=75 mm]{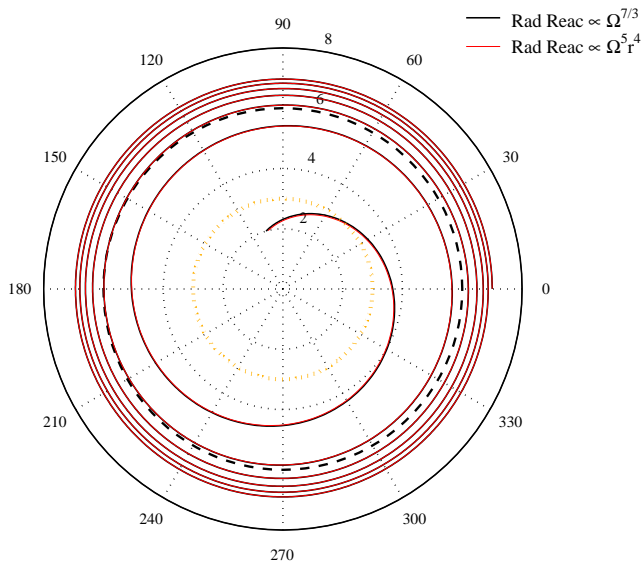}
    \caption{ \label{fig:fig1}Relative dynamics for $\nu =\mu/M=0.01$ and initial separation
     $r=7$. The plot shows the (nearly indistinguishable) trajectories with $\hat{\cal F}_{\varphi}^K$
     (black line) and with $\hat{\cal F}_{\varphi}$ (red line). The dashed circle 
     indicates the LSO (which is crossed at the dynamical time $t/2 \simeq 240$).
      The light ring (at $r=3$), near which the plunge
     waveform will be matched to a ring-down one, is indicated (dotted circle). It is crossed at
     the dynamical time $t/2 \simeq 300$~\cite{Nagar:2006xv}.}
  \end{center}
\end{figure}
Following what we did in~\cite{Nagar:2006xv}, the computation, \`a la 
Regge-Wheeler-Zerilli~\footnote{i.e., by computing the metric perturbations
of a Schwarzschild spacetime by means of a multipolar decomposition. 
See for example~\cite{Nagar:2005ea} and references therein.}
 of the
gravitational waveform in the $\nu\to 0$ limit  needs three separate 
steps: (i) to initialize the 
system~(\ref{eob:1})-(\ref{eob:5}), (ii) to integrate it in time and (iii)
to solve the Regge-Wheeler and Zerilli-Moncrief equations as an initial value
problem, with source terms driven by the particle dynamics. In practice, once
 the dynamics is computed, one needs to solve (for each multipole $(\ell,m)$
of even (e) or odd (o) type)
the following couple of decoupled partial differential equations
\begin{equation}
\label{zm}
\de_t^2\Psi^{(\rm e/o)}_{\ell m}-\de_{r_*}^2\Psi^{(\rm e/o)}_{\ell m} + V^{(\rm
  e/o)}_{\ell}\Psi^{(\rm e/o)}_{\ell m} = S^{(\rm e/o)}_{\ell m} \ 
\end{equation}
with source terms $S^{(\rm e/o)}_{\ell m}$ linked to the dynamics of 
the binary. A thorough discussion of the analytical and numerical details
of our approach has been given in Ref.~\cite{Nagar:2006xv}, so that it is
not necessary to repeat it here. We only recall that the initial condition
for the relative dynamics is given as in~\cite{Nagar:2006xv}, notably by
specifying a non-zero initial value for $p_{r_*}$ (``post-adiabatic'' 
approximation). In addition, the gauge-invariant functions 
$\Psi^{(\rm e/o)}_{\ell m}$ are proportional to the actual
GW polarization amplitudes, and so their knowledge encodes the complete
information about energy, momentum and angular momentum losses.

\subsection{Angular momentum loss}
\label{sbsc:angmom_loss}

Let us consider the relative dynamics of a particle of mass $\mu=0.01M$
plunging from $r=7$. The relative orbit (for both choices of radiation
reaction) is shown in Fig.~\ref{fig:fig1}. The dashed circle indicates the
LSO, which is crossed at $t/2\simeq 240$, while the dotted circle indicates 
the light ring (LR), which is crossed at
$t/2\simeq 300$~\cite{Nagar:2006xv}. 
The first point we want to discuss is the comparison (and contrast) 
between the two expressions of the angular momentum loss mentioned
above, the ``usual'' Keplerian one $\hat{\cal F}_{\varphi}^K$ of 
Eq.~(\ref{FK}) and $\hat{\cal F}_{\varphi}$ of Eq.~(\ref{eq:DGloss}). 
These two expressions are equivalent during the inspiral due 
to the validity of Kepler's law $\Omega^2r^3=1$ along circular orbits,
but they start strongly deviating from each other after the crossing 
of the LSO. We wish to know how they compare to 
some sort of ``exact'' angular momentum flux during the
plunge, and whether one of them is to be preferred.

\begin{figure}[t]
  \begin{center}
   \includegraphics[width=80 mm, height=75 mm]{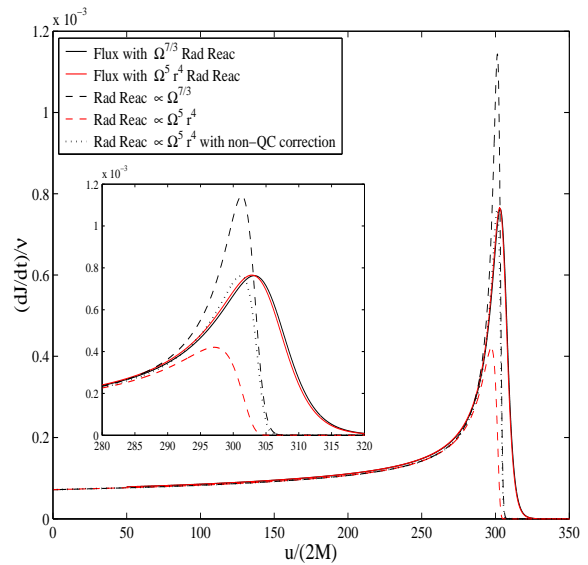}
    \caption{\label{fig:fig2}Comparison between various angular momentum losses:
      GW  fluxes at infinity (solid lines; computed \`a la
      Regge-Wheeler-Zerilli including up to $\ell=4$ radiation multipoles,
      using two types of radiation reaction in the driving dynamics:
      `K' (black) or `not K' (red)); or mechanical angular momentum losses: 
      $(dJ/dt)^{K}=-\hat{\cal F}^K_{\varphi}$ ( hyphenated black lines), 
      $dJ/dt = -\hat{\cal F}_{\varphi}$ (hyphenated red lines).
      The dotted line refers to the mechanical losses with non-quasi-circular corrections.
      See text for discussion.}
  \end{center}
\end{figure}

Let us now be precise in which respect we can compute an ``exact'' result
for the angular momentum loss during the plunge. We recall that via 
Eq.~(17) of Ref.~\cite{Nagar:2006xv} one can compute the instantaneous
angular momentum flux at infinity $dJ/dt$ (as a sum of multipoles) from   
$\Psi^{(\rm e/o)}_{\ell m}$. We have done this for both $\hat{\cal F}^K_{\varphi}$
and $\hat{\cal F}_\varphi$ including all the multipoles of the radiation
up to $\ell=4$. The solid lines in Fig.~\ref{fig:fig2} 
refer to this computation. There, $dJ/dt$ is extracted at $r^{\rm obs}=500$
and is shown versus observer's retarded time \hbox{$u=M(t^{\rm obs}-r_*^{\rm obs})+\Delta u$},
where one shifts the retarded time $u$ by a constant
 so as to best
fit (during the early inspiral) the dynamical time $t$ which enters as the argument
of the local angular momentum loss of the source, i.e. the radiation
reaction  $\hat{\cal F}^K_{\varphi}(t)$ or $\hat{\cal F}_\varphi(t)$.
We use $\Delta u = 0.4 M$.
The labelling ``Flux with $\propto
\Omega^{7/3}$ Rad Reac''  indicates the outcome of 
a dynamics with $\hat{\cal F}^K_{\varphi}$ (black line),
while ``Flux with $\propto \Omega^5 r^4$ Rad reac'' refers to
a dynamics using $\hat{\cal F}_{\varphi}$ (red line). Fig.~\ref{fig:fig2}
shows that the differences in the fluxes at infinity are really tiny. 
In that sense, we know with good accuracy (assuming balance between 
losses at infinity and local losses) what should be the ``exact'' (local) angular momentum 
loss of the system. By contrast, the two dashed lines in Fig.~\ref{fig:fig2}
represent the two analytically-assumed local angular momentum losses; i.e., 
the two radiation reaction terms $\hat{\cal F}^K_{\varphi}(t)$ or $\hat{\cal F}_\varphi(t)$.
The good news is that they agree quite well with the ``exact'' result during
not only the inspiral, but also a fair fraction of the plunge (which
starts at $u/2M \simeq 240$). However, the bad news is that they
start differing significantly from the exact result (as well as from each other)
during the later part of the plunge (say, after $u/2M \simeq 290$,
remembering that the matching to ring-down will be done around $u/2M \simeq 300$,
which corresponds to light-ring crossing). The  ``usual'' Kepler-based
radiation reaction $\hat{\cal F}^K_{\varphi}$ is too large (as anticipated in
Ref.~\cite{Damour:2006tr} which pointed out that $K= \Omega^2 r^3$ sharply
decreases during the plunge, if we notice that the ratio 
$\hat{\cal F}^K_{\varphi}/\hat{\cal F}_{\varphi} \simeq K^{-4/3}$),
while the hopefully better radiation reaction $\hat{\cal F}_{\varphi}$
turns out to be somewhat too small towards the end of the plunge.
On the other hand, the near coincidence of the two corresponding fluxes 
at infinity shows that the ``analytical uncertainty'' in the radiation reaction
does not matter much for computing the waveform. This confirms,
and makes more quantitative, the finding 
of~\cite{Buonanno:2000ef,Buonanno:2000jz} that, after the crossing of the 
LSO, the relative motion is essentially {\it quasi-geodesic}; i.e., 
relatively little influenced by the radiation-reaction force.

The physics behind the differences between the ``analytical'' and ``exact'' angular momentum
fluxes (as well as between the two analytical ones) lies in the presence of 
{\it non-quasi-circular} (NQC) 
effects during the plunge phase. During the adiabatic inspiral one takes
advantage of the quasi-circularity (QC) of the motion to operate {\it two
separate simplifications}: (a) one neglects the terms proportional to 
$dr/dt$ , and (b) one neglects the terms proportional to $d^2r/dt^2$, which,
by using the equations of motion, give (modulo the square of $dr/dt$)
 terms proportional either to $v^2 -1/r$ (``virial combination''), or to
 $\Omega^2 r^3 -1$ (``Kepler combination'' minus 1). During the plunge,
 both simplifications become less and less valid so that one should correct any
 QC-derived expression by terms proportional to powers of the
 two basic dimensionless  NQC ratios $ (dr/dt)/(\Omega r)$ and $(1/r - v^2)/v^2$.
 When considering a time-even NQC correcting factor, it should have the
 general form 
 \begin{equation}
 \label{fNQC}
 f^{\rm NQC}= 1 + a \left(\dfrac{p_{r_*}}{  r \Omega}\right)^2 + a' \left({\dfrac{1}{K}} -1\right) + \cdots  \; ,
 \end{equation}
 where we used the same ``Kepler combination'' $K \equiv \Omega^2 r^3$ as before.
 In our present case, we physically expect the presence of such a time-even
NQC~\footnote{Note that we can henceforth view NQC as meaning either  
{\it non-quasi-circular} or {\it next-to-quasi-circular}.} factor in the
radiation reaction. As we expect (from the usual quadrupole formula, as well
as from the well-known fact that orbital eccentricity tends to {\it increase} 
the flux of angular momentum) that the NQC factor in the radiation
reaction should be {\it larger than one}, we conclude that the ``improved''
radiation reaction expression $\hat{\cal F}_{\varphi}$ proposed 
in~\cite{Damour:2006tr} is a physically more appealing starting point 
for applying such a NQC correcting factor. In addition, as, pragmatically 
speaking, the two types of terms parametrized by $a$ and $a'$ in
Eq.~(\ref{fNQC}) above behave in a roughly similar way during the plunge 
(they are both positive and they both increase), we can try, in first
approximation, to use only one of them, say the first one
parametrized by $a$. 

In Fig.~\ref{fig:fig2} we show (as a black dotted line labelled 
``Rad Reac $\propto \Omega^5r^4$ with non-QC corrections'') the evolution of 
the NQC-corrected radiation reaction
\begin{equation}
\label{eq:F_nqc}
\hat{\cal F}_{\varphi}^{\rm NQC}\equiv 
\hat{\cal F}_{\varphi}\left(1+ a^{RR}\dfrac{p_{r_*}^2}{(r\Omega)^2}\right)
\end{equation}
with the numerical value $a^{RR}= + 2.9$ (fixed by eye). We see how such
a simple (and physically reasonable, because larger than one) NQC
factor suffices to have a good agreement between a ``NQC-corrected analytical
radiation reaction'' and the ``exact''  angular momentum flux $dJ/dt$ up
until the light-ring, that is crossed at $u/(2M)\approx
300$. This gives our first instance of the aim of the present paper: to 
better understand (by comparing to ``exact'' numerical results) the 
``missing physics'' in some first-order analytical expression, and then 
to use this improved understanding to further improve the starting analytical 
expression. Note that, for any starting QC expression $X$ that one wishes to 
correct, it is in principle possible to  compute in General Relativity 
the ``real'' NQC correcting factor 
$f_X^{\rm  NQC}$, of the form~(\ref{fNQC}) with some analytically computable 
coefficients $a_X$, $a'_X$, etc.
Note, however, that one expects these coefficients to become (when considering PN corrections)
functions of $1/r$ and other dynamical variables. As in several other known
cases, one expects that these PN corrections might significantly affect the 
``effective'' numerical value of the coefficients  $a_X$, $a'_X$, etc needed
for a good representation of the considered quantity during the plunge.
It would be interesting to have analytic (at least 1PN-accurate) results on
the coefficients  $a^{RR}_{\cal F}, a'^{RR}_{\cal F}$ to compare them with our
presently ``experimental'' estimate $a^{RR}_{\rm effective}= + 2.9$. 
 
\subsection{Exact waveforms}
\label{sbsc:exact_waves}
\begin{figure*}[t]
  \begin{center}
   \includegraphics[width=75 mm, height=70 mm]{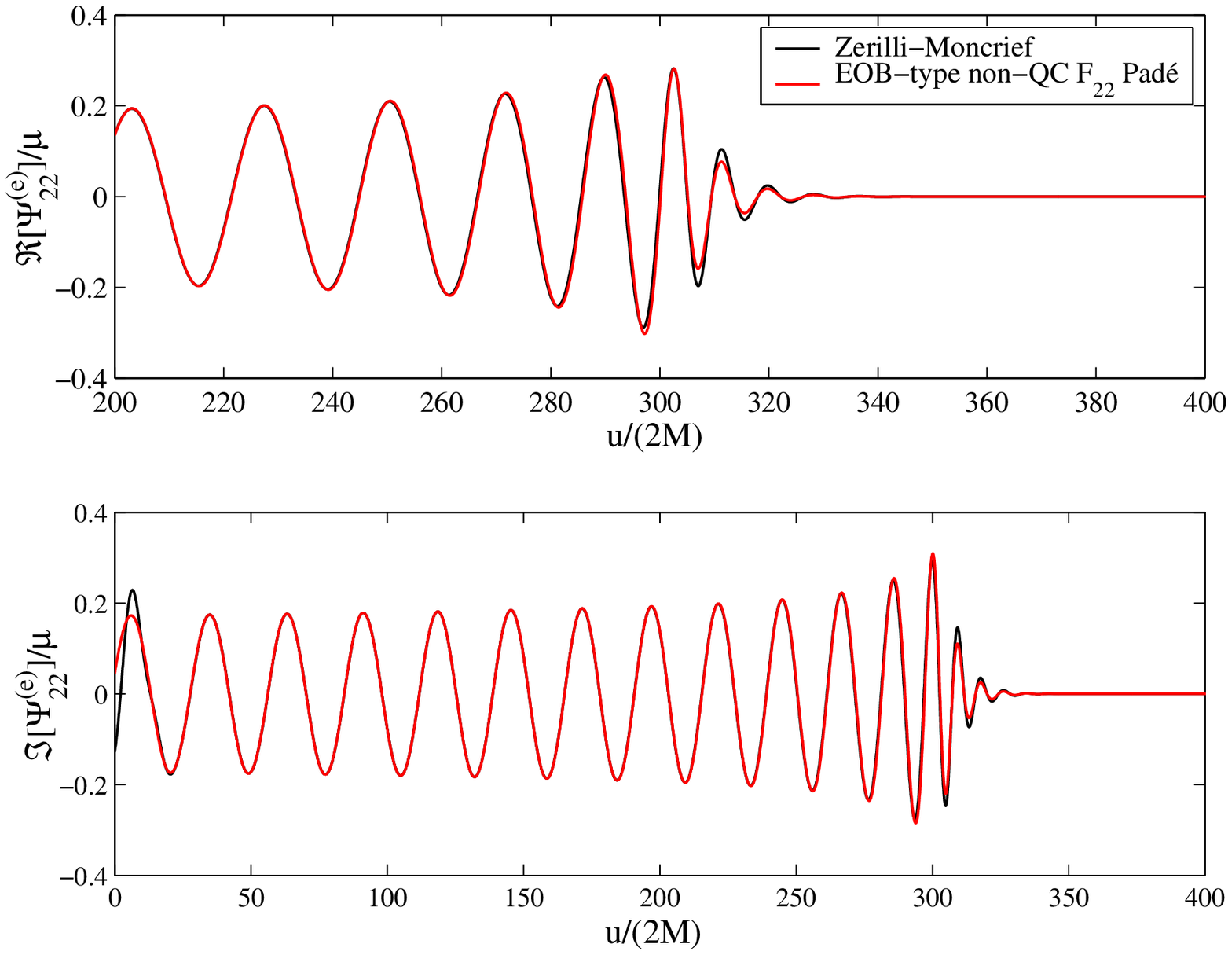} \qquad
   \includegraphics[width=75 mm, height=70 mm]{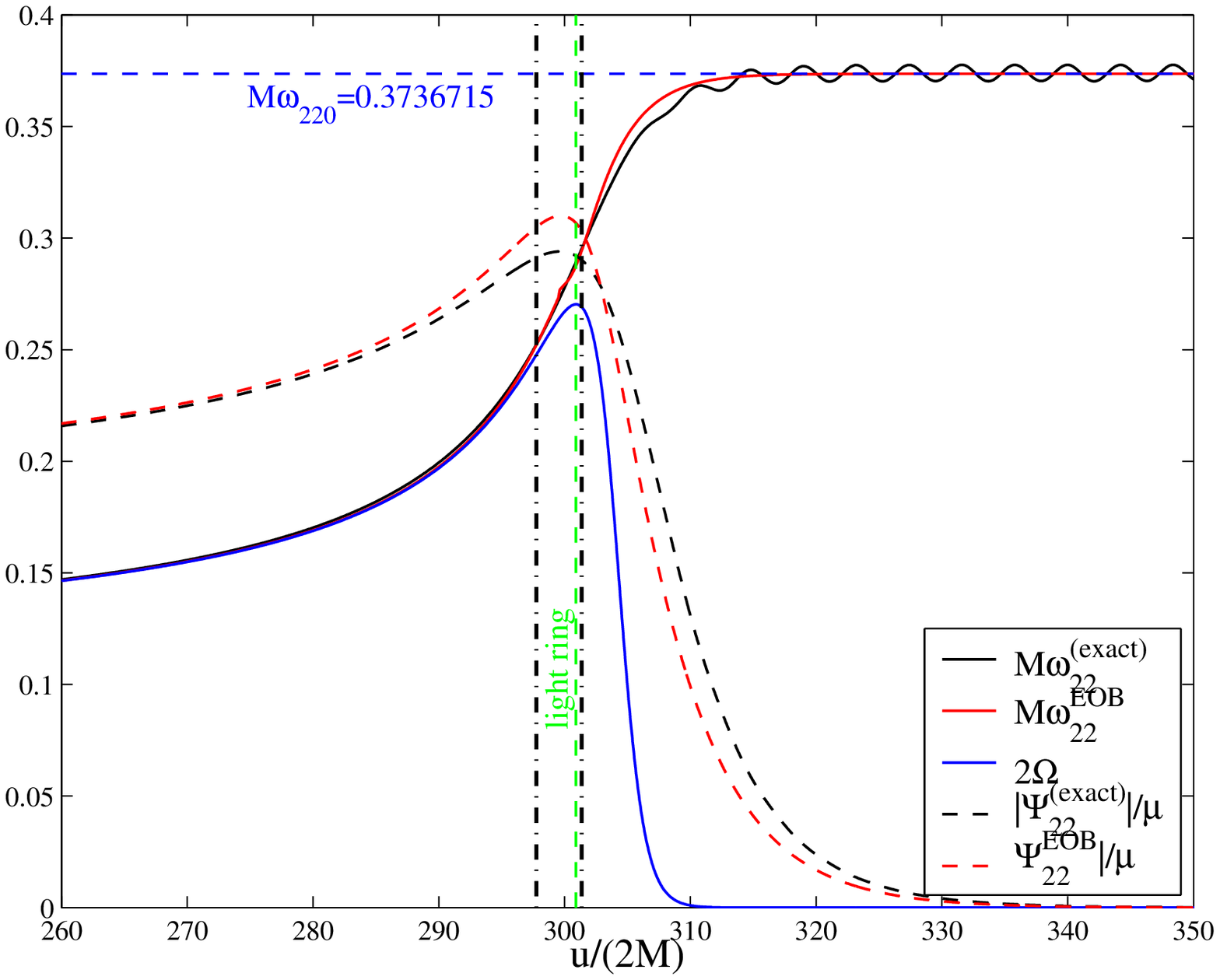} 
    \caption{{\it Left panel}: Quadrupole waveforms ($\ell=m=2$, even parity): 
      exact Zerilli-type (black line) and ``matched'' improved EOB-type (red line).  
      {\it Right panel}: exact and matched-EOB modulus and instantaneous
       gravitational wave frequency. The blue line displays twice the orbital
       frequency $\Omega$. The two vertical dash-dot lines mark the ends (at
       297.8 and 301.4) of our matching interval (which is centered, at 299.6 on the
       maximum of the modulus, and which includes the light ring, at 300.9).
        \label{fig:waveforms}}
  \end{center}
\end{figure*}

Let us now consider the waveforms obtained by numerically integrating the
Regge-Wheeler-Zerilli equation (\ref{zm}), with a source corresponding to the 
plunging dynamics discussed above, with the $\hat{\cal F}_\varphi$ radiation
reaction 
(we have checked that the NQC-improved radiation reaction~(\ref{fNQC}) 
introduces only very minor changes in the waveform).
For the most relevant multipolarity ($\ell=m=2$, with even parity),
the resulting ``exact'' waveforms (denoted $\Psi^{(\rm exact)}_{22}$, or
sometimes $\Psi^{(\rm e)}_{22}$)
are displayed in Fig.~\ref{fig:waveforms}. 
The left panel of the figure 
exhibits (black lines) the real (top) and imaginary (bottom) part of 
$\Psi^{(\rm exact)}_{22}$ (normalized as in~\cite{Nagar:2005ea}).
 The waveforms 
are extracted at $r_{\rm obs}=500$ and are shown versus the (shifted) observer retarded 
time. They exhibit  the usual  ``chirp-like'' structure corresponding to 
the inspiral, and end up with the black hole QNM ringing. 

The most useful information about the dynamics of the merger hidden in
the (exact or analytical) waveform can be extracted by inspection of the
corresponding instantaneous  gravitational wave frequency. 
Since $\Psi_{22}$ is a complex number, this quantity 
is computed as
\begin{equation}
\label{eq:freq_omega22}
M\omega_{22} =\hat{\omega}_{22}\equiv
-\Im \left( \dfrac{\dot{\Psi}_{22}}{\Psi_{22}}\right) \ ,
\end{equation}
where $\Im$ indicates the imaginary part and the overdot stands for $d/dt$. 
The {\it exact} $\hat\omega_{22}$ is shown as a black solid line in the right 
panel of Fig.~\ref{fig:waveforms}. In the same plot we also include, as a blue 
line, twice the orbital frequency $\Omega$ (which is the zeroth-order analytical
prediction coming from any ``standard quadrupole approximation''). 
This plot shows that $\hat{\omega}_{22}^{(\rm exact)}\approx 2\Omega$ during the inspiral 
and most of the plunge
(the LSO is crossed at $u/(2M)\simeq 240$). 
It is only  when the (relative) radius gets quite near the light-ring 
value $r=3$ (at $u/(2M)\approx 300$, indicated by a green dashed 
vertical line) that the two frequencies start deviating from each other. 
While the doubled orbital frequency  $2\Omega$ reaches a maximum 
equal to $ 2 \Omega_{\rm max} = 0.2703$ very near
the light ring~\footnote{It is easy to see from the Hamiltonian
equation (\ref{eob:3}) why this is so: neglecting the relatively
slow variations of the angular momentum and the energy, $\Omega$ vary
proportionally to the ratio $A(r)/r^2$ which is precisely the effective radial
potential for photon (or ultra-relativistic) circular orbits.} 
and then exponentially decreases, the exact GW frequency 
$\hat{\omega}_{22}$ keeps growing after the light ring
 until it saturates and oscillates around the fundamental 
QNM frequency of the black hole, 
$M\omega_{220}=0.3736715$~\cite{Chandra:1975,Leaver:1985ax}.
The  oscillation in the GW frequency around the fundamental QNM
frequency is (essentially) due to
the superposition of both {\it positive-}~\footnote{As in Quantum Field Theory we call
{\it positive} frequency a mode $\propto e^{- {\rm i} \omega t}$ with $\omega >0$. We also
use for the QNMs  (see below) the notation $\propto e^{-  \sigma t}$. In this notation,
the positive frequency fundamental mode has  $\sigma^+_0=\alpha_0+{\rm i}\omega_0$, 
where $\omega_0$ indicates the frequency and $\alpha_0$ the damping time of
the fundamental mode, while the negative frequency one has
$\sigma^-_0=\alpha_0-{\rm i}\omega_0$.} and negative-frequency 
excitations in the ring-down waveform.~\footnote{Evidently, had we initialized the dynamics 
with the opposite sign of $p_{\varphi}$, i.e., the particle inspiralling 
{\it clockwise} and not {\it counterclockwise}, we would have obtained a 
mirror-reflected result, with the negative frequency modes being more excited than the 
positive frequency ones.}
Let us mention that this kind of behavior of 
$M\omega_{\ell  m}^{(\rm e/o)}$ (odd and even-parity GW frequencies)
is absolutely general and the amplitude of the oscillation depends
on $m$~\cite{DN07c}.

The fact that the {\it amplitude} of the frequency oscillation in the right 
panel of Fig.~\ref{fig:waveforms} is quite small then means that the 
positive-frequency mode is excited with a much larger coefficient than the negative-frequency one.
In view of our general aim of understanding all the separate physical elements 
crucially entering the final waveform, it is interesting to explain what is the basic
underlying physics behind this large dissymmetry in the excitation coefficients 
of the positive-frequency and negative-frequency modes (we have checked that this
dissymmetry extends to the first few overtones above the fundamental QNM mode).
Though this basic physics can be described in terms of equations, it is best
understood by means of some diagrams. One should have in mind that solving the
Regge-Wheeler-Zerilli equation (\ref{zm}) means essentially considering a
point-source (moving along the radial axis) whose ``strength'' oscillates  proportionally to 
$\exp[- {\rm i} m \varphi(t)]$ (with $m= + 2$ in the present case) and which is located
at a time-dependent radius $r_*(t)$. If the radial velocity is small with respect to 
the azimuthal velocity (as is true during most of the plunge) we can think of a source
with an adiabatically varying frequency $m \Omega(t)$ located at the 
adiabatically varying radius $r_*(t)$. Replacing this behavior in equation (\ref{zm})
allows one to approximately replace the repeated time derivative by the square
of the instantaneous source frequency, i.e. by a factor 
$(- {\rm i} m \Omega(t))^2 = - (m \Omega(t))^2$.
This yields a Schroedinger-type (radial) equation with effective potential
$V^{(\rm e/o)}_{\ell}(r_*) - (m \Omega(t))^2$.
\begin{figure}
\includegraphics[width=85 mm, height=75 mm]{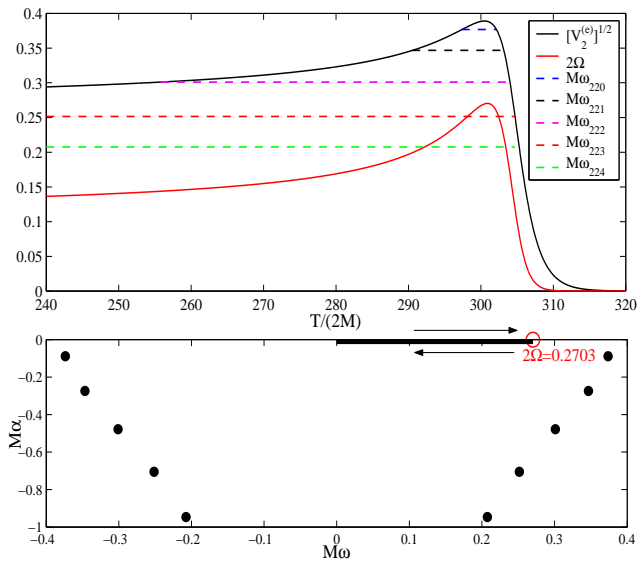}
\caption{\label{fig:fig4}Illustration of some of the important physical 
features of the excitation of quasi-normal modes by a small-mass-ratio 
coalescing binary system. The upper panel shows that, during all the plunge, 
one remains in a ``quasi-stationary'' near-zone regime (which does not 
excite QNMs). The crucial feature which can excite the QNMs is the 
{\it non-adiabaticity} of the evolution of the exciting orbital frequency 
near its maximum (similarly to Ref.~\cite{Damour:2006tr})\textbf{}.
The lower panel illustrates why the (rather short) non-adiabatic behavior 
of $2 \Omega(t)$ near its maximum preferentially excites the 
positive-frequency QNMs.
} 
\end{figure}
In the upper panel of Fig.~\ref{fig:fig4}, we compare the
evolution, as a function of dynamical time $t$, of the square roots
of the two contributions to this effective radial potential: $\sqrt{V^{(\rm e/o)}_{\ell}(r_*(t))}$
and $m \Omega(t)$ ( for the case $\ell = m =2$). We see that the $(m \Omega(t))^2$ term
is always significantly smaller than the $V^{(\rm e/o)}_{\ell}(r_*(t))$ one. This means
that during all the plunge (including at the end, when $\Omega(t)$
reaches its maximum value), we can think (in first order) in terms of a {\it near-zone} approximation,
where we can neglect $(2 \Omega(t))^2$ and compute the asymptotic waveform by convolving the
source by a solution of the static, radial Regge-Wheeler-Zerilli equation. One can check that this
leads essentially to the usual (Schwarzschild-coordinate) quadrupole formula for
the waveform. In this approximation there is strictly no excitation of the QNM modes
by the adiabatically varying source. One therefore concludes (following the line
of reasoning of Ref.~\cite{Damour:2006tr} where it was explained how a certain
type of oscillating integral which exponentially vanishes in the adiabatic approximation
acquired a non-zero value due to a localized non-adiabatic behavior of the
integrand) that the QNM modes will be excited essentially only around the
maximum of the exciting frequency $m \Omega(t)$, because this is the moment
when the source excitation is {\it non-adiabatic}. Looking now at the bottom panel
of Fig.~\ref{fig:fig4}, we see the trajectory of the exciting frequency on the
{\it positive} real axis, from zero up to the maximum $2 \Omega_{\rm max}
= 0.2703$. The black dots in the complex frequency plane represent the
first QNM modes of a Schwarzschild black hole. One understands intuitively
(and this can be put in equations similar to those derived in~\cite{Damour:2006tr})
that the amount of excitation, by the non-adiabatic behavior of the excitation
frequency near its maximum, of the various QNM modes will primarily
depend on their ``distance'' to the critical (real) exciting frequency 
$+ 2 \Omega_{\rm max} =  + 0.2703$. This reasoning indicates that the QNM
modes which will be primarily excited are the first few {\it positive frequency}
modes, i.e. those appearing on the right part of the bottom panel of 
Fig.~\ref{fig:fig4}. Summarizing in slightly different physical terms, 
we can think of the black hole as a resonating object having the resonance 
spectrum showed in Fig.~\ref{fig:fig4}. When the source varies adiabatically, 
so that it corresponds to a nearly pure frequency $ + m \Omega$ which is 
much {\it smaller } than any of the resonance frequencies it excites only 
negligibly the resonance modes. However, when the source varies 
non-adiabatically (near its maximum frequency), its  Fourier transform 
contains a halo of frequencies (due to non-adiabaticity) centered around the
instantaneous frequency, and this halo can now ``extend the reach'' of 
$+ m \Omega_{\rm max}$ and thereby excite the black hole resonances 
which lie within this ``halo'' (i.e. the first few QNM modes whose 
frequency has the {\it same sign} as $+ m \Omega_{\rm max}$). 

\section{EOB-type approximate waveforms}
\label{sec:eob_waves}

\subsection{Methodology: a matched waveform}
\label{sbsc:match_wave}
Let us now turn to the construction of complete EOB-type gravitational
waveforms obtained by {\it matching} an analytical {\it inspiral-plus-plunge} waveform
to  a {\it ring-down} one. For the reasons explained above, the crossing of 
the light ring (actually the non adiabatic behavior around the
moment when $\Omega(t)$ reaches it maximum) corresponds to an abrupt
triggering of the QNMs of the black hole (this was realized long ago~\cite{Davis:1971gg} for the case of the radial plunge).
The EOB approach takes into account this abrupt change in the underlying physics
by an abrupt change in the analytical description of the system:
before the crossing of the light ring one describes the binary system as two point masses
with EOB relative dynamics, while after the crossing one replaces the
binary system by a single distorted black hole (as in the ``close-limit''
approximation of colliding black holes~\cite{Price:1994pm}). As we were
mentioning in the introduction, at the practical level this dual
description implies that one should {\it match} together (around the light ring) 
a plunge waveform obtained with a {\it quadrupole-type} formula to a superposition
of black hole QNMs. This general philosophy has already been implemented, with
increasing sophistication in Refs. \cite{Buonanno:2000ef,Buonanno:2005xu,Damour:2006tr,Buonanno:2006ui,Pan:2007nw}.
Our aim here is to further improve the implementation of the EOB approach by studying 
in detail each one of its building blocks, and by comparing the resulting analytical waveform
to the {\it exact} numerical waveforms discussed above.

\subsection{An improved, resummed 3PN-accurate inspiral-plus-plunge waveform.}

We start by introducing here a new, improved EOB-type inspiral-plus-plunge waveform that goes beyond 
previous work in two directions: (i) it includes higher post-Newtonian (PN) effects
which are {\it resummed} in a novel way (factorized formulation, exact resummation of
tail effects, 3PN-resummed non-linear PN effects), and 
(ii) it includes non-quasi-circular (NQC)
corrections that are expected to be relevant below the LSO. Details
of our construction will be explained separately \cite{DN07c}.
Up until the light ring $(r\gtrsim 3$), we use a $(\ell,m)= (2,2)$
 gravitational waveform defined by means of the following 
 EOB-type improved quadrupole formula 
\begin{align}
\label{eq:psiEOB_inspiral}
\Psi^{\rm plunge}_{22}(t) &= -4\mu\sqrt{\dfrac{\pi}{30}} \left(r\Omega\right)^2 F_{22}\nonumber\\
&\times\left\{1+a\dfrac{p_{r_*}^2}{(r\Omega)^2} + {\rm
  i}b\dfrac{p_{r_*}}{r\Omega}\right\}e^{-2{\rm i}\varphi} \ .
\end{align}
Here, $F_{22}$ includes four types of PN effects, while the braces represent
NQC corrections. The standard quadrupole formula (for a point-particle source
in quasi-circular motion) corresponds to
the case where $F_{22}=1$ and $a=b=0$  (see e.g.~\cite{Thorne:1980ru}; the
numerical coefficient $\sqrt{\pi/30}$ corresponds to our using the 
Zerilli-Moncrief even quantity $\Psi^{(\rm e)}_{22}$).

 The NQC factor added above (within braces) is more general than the general {\it time-even}
 NQC factor considered in (\ref{fNQC}) because we describe here  non time-even
 NQC effects (as can easily be seen on the usual quadrupole formula written
 with two time-derivatives). The numerical coefficients $a$ and $b$ thus
parametrize  non-quasi-circular corrections to the modulus and to the 
phase of the wave. They will be fixed below by comparison to the ``exact'' 
Zerilli-Moncrief waveform. 
The novel $F_{22}$ PN-improving factor is given as the  product of four terms
\begin{equation}
\label{eq:F22}
F_{22}(t)=\hat{H}(t)T(2,2\Omega(t)) e^{{\rm i}\delta_{22}(t)} f_{22}(x(t))\ .
\end{equation}
This factorization of PN effects is, in itself,  a new way of partially resumming
PN effects (by contrast to usual representations of PN effects which
mix the expansion of the four separate factors above). 
The first factor is the ``conserved'' relativistic energy of the (effective) moving source.
The second factor 
$T(2,2\Omega)$ is the particularization to $\ell=m=2$ of a  
function $T(\ell,k)$ (with $M k\equiv \hat{k} = m  \Omega$) that represents a ``tail'' correction, 
namely the contribution due to backscattering of (multipolar) 
waves off the background curvature. This factor is the exact resummation of an 
infinite number of ``leading logarithms'', and is derived, similarly to the
work in Ref.~\cite{Asada:1997zu}, by using Coulomb wave functions.
It is explicitly given by 
\begin{equation}
\label{eq:tail_term}
T(\ell,k) = \dfrac{\Gamma(\ell+1-2{\rm i}\hat{k})}{\Gamma(\ell +1)}e^{\pi \hat{k}} e^{2{\rm i}\hat{k}\log(2 k b_{\rm scale})} \ ,
\end{equation}
where we shall take $b_{\rm scale}=2M$.  $e^{{\rm i}\delta_{22}}$ is an additional phase correction, with
$\delta_{22}=7  \Omega/3$ to lowest PN order. The final factor  $f_{22}(x)$ represents the remaining
(essentially non-linear) PN effects. Using Ref.~\cite{Tagoshi:1994sm} it can be computed in the
test-mass limit, during the inspiral,  as a 
Taylor expansion in the PN-ordering parameter $x = 1/r = \Omega^{2/3}$. At 
3PN~\footnote{We note that for $\nu\ll 1$
this series is known to  higher PN orders. We truncate it here to the 3PN level
which is the highest order at which we could (and did) derive it
 in the comparable-mass case. We leave to a future publication a detailed
 discussion of the derivation of these results.}, it reads 
\begin{align}
\label{eq:f22}
f_{22}^{\rm Taylor}(x)&=1-\dfrac{43}{21}x-\dfrac{536}{189}x^2 +
x^3\bigg[\dfrac{21428357}{727650}\nonumber\\
             &-\dfrac{856}{105}\gamma -\dfrac{1712}{105}\log(2) -
             \dfrac{428}{105}\log(x)\bigg] \ ,
\end{align} 
where $\gamma=0.577216\dots$ is the Euler constant. 
To improve its convergence properties in the strong-field-fast-motion 
regime, expression~(\ref{eq:f22}) is further Pad\'e resummed in a 
proper way. 
 
 When deriving the factor $f_{22}(x)$ during the inspiral phase, the
 argument $x$ that it depends upon is ``degenerate'' in the sense that
 it can be expressed in various ways in terms of the dynamical variables
 of the system, namely: $x=v_{\varphi}^2= (r \Omega)^2= \Omega^{2/3}=1/r$.
 However, these various expressions start differing from each other during the
 plunge. This opens the issue of choosing the ``best'' dynamical expression for
 the PN-ordering argument $x$ during the plunge, in the sense of giving
 the best approximation to the 
``exact'' waveform (minimizing the need for NQC corrections). 
We have found that choosing $x=1/r$ (the Newtonian
gravitational potential) yields good results, and this is the choice
we shall make here. We postpone to a
further publication a thorough analysis of the effect of different
$x$ arguments, as well as the contrast between the Pad\'e-resummed 
(that we shall use here) and Taylor-expanded versions of $f_{22}(x)$.

\subsection{Ring-down waveform}

After the crossing of the light ring, the EOB approach consists in replacing
the two-body system by  a single distorted black hole. 
At the level of waveforms, this means that one should {\it match}, around
the light ring, the plunge waveform Eq.~(\ref{eq:psiEOB_inspiral}) 
to a {\it ring-down} one made of a superposition of ($\ell=2, m=2$) 
black hole QNMs, say
\begin{equation}
\label{eq:psiEOB_ringdown}
\Psi^{\rm ringdown}_{22}(t) = \sum_n C_n^{+} e^{-\sigma_n^{+} t}+ \sum_n C_n^{-} e^{-\sigma_n^- t} \ ,
\end{equation}
where $\sigma_n^{\pm} = \alpha_n\pm{\rm i}\omega_n$ are the 
positive/negative  complex QNM frequencies and $C_n^{\pm}$ are the corresponding amplitudes. 
Here, $\alpha_n$ and $\omega_n$  indicate the damping time and the 
oscillation frequency of each mode respectively, and $n=0,1,2,...,N-1$ labels the 
overtone number ($n=0$ denoting the fundamental mode). 

In this paper we propose an improved treatment of the definition of the
ring-down waveform Eq.~(\ref{eq:psiEOB_ringdown}), as well as of the matching
to the plunge waveform Eq.~(\ref{eq:psiEOB_inspiral}). First, in view of both
the numerical results and the theoretical arguments presented above, we know
that the contribution of the positive-frequency modes to the waveform
is much larger than that of the negative-frequency ones. Therefore, we can set $C_{n}^-=0$
in first approximation. As a result, the instantaneous 
gravitational wave frequency $M\omega_{22}^{\rm EOB}\equiv
-\Im(\dot{\Psi}^{\rm EOB}_{22}/\Psi^{\rm EOB}_{22})$ of our ``matched'' ring-down waveform
will grow until it reaches a ``flat'' saturation region, {\it without oscillations}.
Though this is qualitatively slightly different from the behavior of the exact
instantaneous frequency (see Fig.~\ref{fig:waveforms}) we will see that 
this gives an excellent approximation to the phasing (and actually a 
better one that if one matches to both positive and negative modes in 
a democratic way).

Our problem is thus to determine the coefficients $C_n^{+}$ from the 
knowledge of $\Psi_{22}^{\rm plunge}$ for $r\gtrsim3$. Here again we introduce
some significant improvements on previous implementations of the EOB philosophy,
based on the knowledge acquired from detailed comparisons with our ``exact''
small-mass-ratio numerical results. A trivial improvement will consist
in using a larger number $N$ of QNMs (including the fundamental mode) than previous work. 
Instead of $N=1$~\cite{Buonanno:2000ef,Buonanno:2005xu}, 
$N=2$~\cite{Damour:2006tr}, or $N=3$~\cite{Buonanno:2006ui,Pan:2007nw},
we shall use here $N=5$ (positive-frequency) quasi-normal modes 
(including the fundamental mode)~\cite{Chandra:1975,Leaver:1985ax}.
Our studies showed that $N=5$ gives a significantly better fit than $N=4$ or less,
and that larger values of $N$ yield only a rather marginal improvement.

\subsection{Determining the next-to-quasi-circular (NQC) corrections to the plunge waveform.}

 First of all $\Psi_{22}^{\rm plunge}$, Eq.~(\ref{eq:psiEOB_inspiral}),
has two free parameters $a$ and $b$ that should be specified so to minimize 
the difference with the real signal. The presence of two separate NQC real
parameters is useful because the waveform is a complex number, and we wish to
best fit both its modulus and its phase\footnote{In view of this, a useful alternative
to the  $a,b$ NQC factor in braces used above, is to use instead a factorized complex NQC
waveform correction of the form, $ (1+a' p_{r_*}^2/(r\Omega)^2))\exp (+{\rm i}b' p_{r_*}/(r\Omega))$,
in which $a'$ affects only the modulus, and $b'$ only the phase.
The (approximate) analytic constraint on the modulus discussed below then becomes simply $ 2 a'=1$.
}. The {\it modulus} of the numerical waveform reaches
a maximum at the (shifted) retarded time $u/(2M)= 299.6$ (i.e. just before the 
maximum of the orbital frequency, or the crossing
of  the light ring, which both occur near $u/(2M)=
300.9$). 
We can then constrain the NQC parameters
 $a$ and $b$ so as to ensure that the maximum of the modulus of the 
 analytical waveform (\ref{eq:psiEOB_inspiral}) sits on top of that of 
the exact modulus.  Actually, we can even do this
{\it analytically}. Using the quasi-geodesic approximate description of the plunge,
and imposing that the modulus of the analytical waveform (\ref{eq:psiEOB_inspiral})
is maximum at the light-ring gives, in first NQC approximation, the constraint 
$2 a + b^2 =1$. We then get a second constraint (mainly on the coefficient $b$) by
requiring that the {\it slope} of the instantaneous analytical frequency 
$M\omega_{22}^{\rm plunge}(t)$ be as close as possible to the 
exact one $M\omega_{22}(t)$ around the light ring. The two above requirements
fix the values of the NQC parameters to be: $a=0.438$ and $b=0.355$. It is remarkable
that these two numerical values (numerically determined, by trial and error,
 as best eye-fitting values)
do satisfy quite accurately the (first-order) analytically
determined constraint mentioned above: indeed, $2 a_{\rm num} + b_{\rm num}^2=1.002$.
This shows the power of the EOB approach for 
building a purely analytical {\it self-consistent} matched waveform, 
even in absence of knowledge about the exact signal.

\subsection{Matching the inspiral-plus-plunge waveform to the ring-down one on a ``comb''.}

The last and most crucial new ingredient introduced here consists in
{\it matching} the plunge and ring-down waveforms 
on a {\it finite time interval} $[t_m - \delta/2,t_m +\delta/2]$, 
instead of simply imposing the continuity of the two functions, 
and a certain number of their derivatives,  at one moment 
$t_{\rm match}$ 
(as done in~\cite{Buonanno:2000ef,Damour:2006tr,Buonanno:2006ui,Pan:2007nw}). 
We found that this matching  on a finite interval  (chosen in the immediate vicinity of
light-ring crossing) was quite effective for accurately reproducing the
smooth rise of the instantaneous GW frequency during the whole transition.  
More precisely, a good choice for the matching interval $ [t_m - \delta/2,t_m +\delta/2]$
 consists in putting its center $t_m$ at the {\it maximum of the 
modulus} of  $\Psi^{\rm plunge}_{22}$, i.e. at $t/(2M)= 299.6$, and choosing 
his full length  $\delta = \Delta t$ as $\delta /M\simeq 7.2$. The matching 
on the interval $ [t_m - \delta/2,t_m +\delta/2]$ is most easily done by 
discretizing this matching interval into $N$ (here $N=5$) equally spaced
points, $[t_1,t_2,\ldots,t_N]$ (with $t_1=t_m - \delta/2$ and $t_N=t_m +\delta/2$)
 and requiring that the two complex functions 
(\ref{eq:psiEOB_inspiral}) and (\ref{eq:psiEOB_ringdown}) 
coincide at these $N$ points. This gives $N$ (linear) complex equations 
for the $N$ complex unknowns $C_n^{+}$. In other words we are using a `matching comb'.

\subsection{Comparing analytical and numerical waveforms}

Finally, we define our complete 
EOB matched waveform (from inspiral to ring-down) as
\begin{align}
\label{eq:psiEOB_matched}
\Psi^{\rm EOB}_{22}(t)&\equiv \theta(t_m -t) \Psi^{\rm plunge}_{22}(t)\nonumber\\
                      &+ \theta(t -t_m ) \Psi^{\rm ringdown}_{22}(t)
\end{align}
where $\theta(t)$ denotes Heaviside's step function\footnote{If one wanted to have a
$C^{\infty}$ transition between the two waveforms one could replace $\theta(t -t_m)$ by one
of Laurent Schwartz's well-known smoothed step functions (or ``partitions of unity'')
 ${\theta_{\epsilon}}((t-t_m)/\delta)$}.

The left panel of Fig.~\ref{fig:waveforms} compares the (real and imaginary
parts of the) matched  EOB-type waveform (\ref{eq:psiEOB_matched})(red line)  
obtained by means of the matching procedure explained above, to the ``exact'' 
numerical one (black line). The agreement between the two is clearly
excellent. Note, in particular, how the inclusion of the novel 3PN-accurate
correction factor $F_{22}$ allows the analytical {\it inspiral} waveform
to coincide in phase {\it and} in amplitude with the exact signal.
As we shall illustrate below, Newtonian-accurate ``restricted'' waveforms 
do a much poorer job: though they describe the phasing rather accurately,
they overestimate (already during the inspiral) the amplitude by a significant
factor.
To investigate further the quality of this agreement, we also plot
the modulus of the waveforms and  the instantaneous frequency (right panel
of Fig.~\ref{fig:waveforms}). One sees again on this figure the very accurate way
in which the improved EOB signal reproduces both the frequency and the modulus
of the exact signal during the inspiral and most of the plunge (which starts at
$u/2M \simeq 240$). It is only at the end of the plunge,
just before light-ring crossing, (and during ring-down)
that one notices a difference (smaller than about $5\%$) in modulus. 
\begin{figure}[t]
  \begin{center}
   \includegraphics[width=75 mm, height=70 mm]{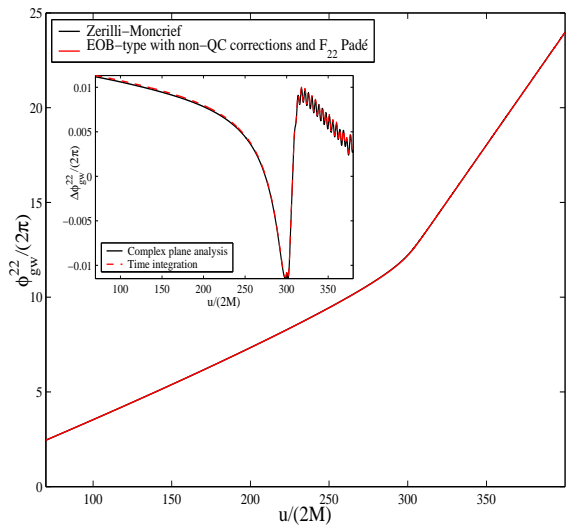}
    \caption{\label{fig_frequency}Analytic and numerical 
      gravitational wave phases versus (shifted) retarded time.
     As they are very nearly superposed, the inset shows the difference between the two
     phases (computed, as a check, by means of two independent methods).}
  \end{center}
\end{figure}
To further study the difference in frequency of the two signals,
we compare in Fig.~\ref{fig_frequency} the time evolution of the {\it phases}
of the two signals (with the convention $\Psi = e^{ - {\rm i} \phi}$, so that
$\dot{\phi} = + \omega$):
 the $matched$ instantaneous gravitational 
wave phase $\phi_{22}^{\rm EOB}$ versus the exact one $\phi_{22}^{(\rm exact)}$.
 These quantities are computed from 
time integration of the instantaneous frequencies from an initial
time $t_0$ starting with $\phi_0 = 2\varphi_0$ where $\varphi_0$ is the initial orbital
phase. The visual agreement between the exact phase (black line) and the 
matched phase (red line) is {\it so good that the two lines can barely be 
distinguished}. At a quantitative level, the phase difference
$\Delta \phi^{\rm gw}_{22}\equiv\phi^{\rm EOB}_{22}-\phi^{(\rm exact)}_{22}$ 
is an important diagnostics which is shown in the inset of of 
Fig.~\ref{fig_frequency} (after adding a constant offset). This has been 
computed by two separate methods. The first one (red dashed line) computes
$\Delta \phi^{\rm gw}_{22}$ by
 straightforward time integration of the difference between
$M\omega^{\rm EOB}_{22}$ and $M\omega_{22}$. 
The second computation (black solid line) does not involve any time integration,
it just evaluates the argument of the ratio of the two
complex numbers $\Psi^{\rm EOB}_{22}$ and $\Psi^{({\rm exact})}_{22}$:
\begin{equation}
e^{-{\rm i}\Delta\phi_{22}^{\rm gw} } = \dfrac{\Psi^{\rm EOB}_{22}}{\Psi^{(\rm exact)}_{22}}
\dfrac{|\Psi^{(\rm exact)}_{22}|}{|\Psi^{\rm EOB}_{22}|} \ .
\end{equation}
As shown in the inset of Fig.~\ref{fig_frequency} there is an excellent consistency between the
two methods. 

Fig.~\ref{fig_frequency} exhibits our most important final result: the maximum phase difference, 
during the whole transition from inspiral to plunge,
between our improved EOB signal and the exact one is of the order of $\pm 1.1\%$ of a cycle. 
Note that the \textit{}ripples in the late-time part of $\Delta\phi_{22}^{\rm gw}$ are
(obviously) related to the absence of negative frequency modes in the
matched waveform. 

This excellent analytical description of the exact waveform (phasing to $\pm
1.1\%$ of a cycle, and a rather accurate modulus, see Fig.~\ref{fig:waveforms})
was obtained by combining the several improvements  to the EOB method introduced
above, and by using the specific numerical values of the new parameters 
$a_{RR}$, $a$, $b$, $t_m$, $\delta$ quoted above. 
Note, however, that there remains some flexibility in the numerical values 
of these parameters. We leave to future work a detailed discussion of this 
issue. Let us only note here that the ``good choices'' of these parameters 
(minimizing the maximum (absolute) dephasing~\footnote{We shall refer to 
the minimum value of the maximum absolute value of the dephasing 
(over the whole inspiral-plunge-merger period) as the ``minimax'' dephasing.}) 
are correlated, and that the values we have used above are near optimal
in the following sense: if, for example, we keep fixed the matching 
comb parameters $t_m$ and $\delta$, and vary $a$ in the 
interval $[0.3,\,0.8]$, the best possible complementary values of $b$ 
that we could find (varying in the interval $[0.2,\, 0.5]$) were found 
to lead to minimax dephasings larger than $1.1\%$ ,
but remaining smaller than $3\%$. We also explored the sensitivity 
to the choice of the matching comb. For instance, if we keep fixed 
only the center of the matching interval at $t_m/(2M)=299.6$, and increase 
$\delta/M$ by a factor 2 (to $\delta/M = 14.4$) and then look for the best 
possible values of the parameters $a$ and $b$, we find that the minimax 
dephasings are increased from $1.1\%$ to about $2\%$. 
On the other hand, if (still keeping fixed $t_m$) we decrease $\delta/M$ to
$1.8$, and then look for the best possible $a$ and $b$,
we find that the minimax dephasings are increased from $1.1\%$ to $2.5\%$.
Similar sensitivities are also reached if we keep fixed $\delta/M$ 
(to the near optimal value $7.2$), but move the center of the interval: 
for example, $t_m/(2M)= 298$ leads to a minimax of $2\%$, 
while $t_m/(2M)=301$ gives $4\%$, so that we obtain correspondingly 
less accurate decriptions of the waveform.
 
On the other hand, as we shall discuss in detail in a separate publication, even much 
simplified versions of our matching technique can lead to  rather 
impressively good results, notably for the phasing. To quote one 
particularly simple example, the standard quadrupole formula in the 
quasi-circular approximation (in the sense of Eq.~(\ref{eq:psiEOB_inspiral}) 
with $F_{22}=1$, $a=b=0$ and $N=5$), still gives a quite tiny dephasing, 
$\Delta\phi_{22}^{\rm gw}/(2\pi)\lesssim 2\%$, though the evolution 
of the modulus is less well reproduced.

\begin{figure}
\includegraphics[width=75 mm, height=70 mm]{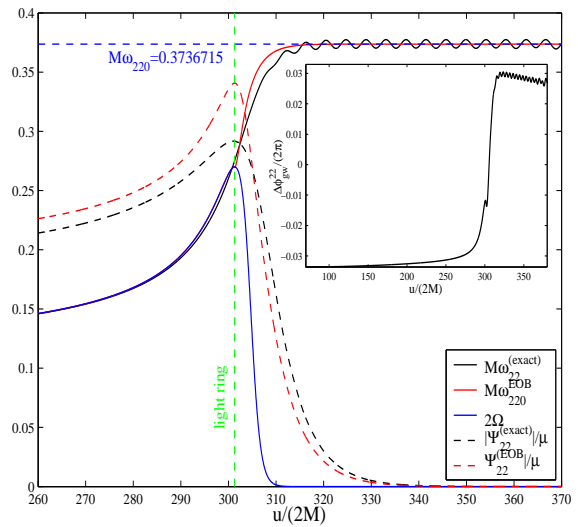}
\caption{\label{fig:fig6}Comparison between the moduli and the phases of 
two waveforms: the ``exact'' numerical one, and a non-optimized analytic 
one based on a coarser implementation of the EOB philosophy (no PN corrections,
no explicit NQC corrections, $N=3$ QNMs, nearly instantaneous matching at the
light ring). 
See text for details.}
\end{figure}
Finally, we pedagogically illustrate in Fig.~\ref{fig:fig6}, in a ``contrasting'' manner, 
the effect of the various improvements introduced above by showing how the 
waveform comparison of Fig.~\ref{fig:waveforms} gets modified when 
{\it not} using them, but using instead the type of coarser implementation 
of the EOB philosophy used in previous 
work~\cite{Buonanno:2000ef,Buonanno:2005xu,Damour:2006tr,Buonanno:2006ui,Pan:2007nw}.
More precisely, the EOB-type analytical waveform used in Fig.~\ref{fig:fig6} 
was obtained by: (i) using (as originally proposed in Ref.~\cite{Buonanno:2000ef}) 
the following (Newtonian-order and Kepler-law-assuming) restricted quadrupole 
waveform 
\begin{equation}
\label{eq:psiEOB_BD}
\Psi^{NK}_{22}(t) = -4\mu\sqrt{\dfrac{\pi}{30}} \Omega^{2/3} e^{-2{\rm i}\varphi} \ ,
\end{equation}
without any explicit  PN  ($F_{22}$) corrections, nor any NQC ($a,b$) corrections;
(ii) only $3$ (positive-frequency) QNMs; and, rather crucially, (iii) by matching the plunge
and ring-down waveforms in a very small interval ($\delta/M = 0.9$ instead of our
preferred $7.2$) around the maximum of the orbital frequency. [Indeed, the 
 matching of the two waveforms and their derivatives at a sharply defined moment
is equivalent to considering the $\delta \to 0$ limit of our 
comb-matching technique]. By contrasting Fig.~\ref{fig:fig6} with the 
right panel of Fig.~\ref{fig:waveforms} we see that: 
(i) the modulus of the analytical waveform is now distinctly
larger than the exact one during the inspiral (because of the lack of PN corrections);
(ii) the modulus becomes significantly larger than the exact one at the end of the
plunge (because of the use of the Kepler-law-assuming $\propto \Omega^{2/3}$, which,
as pointed out in~\cite{Damour:2006tr} tends to
overestimate the amplitude); (iii) the post-matching analytical frequency jumps up from the
(doubled) maximum orbital frequency significantly more vertically than before,
thereby decoupling too soon from the exact frequency, and accruing a larger dephasing
than before (because of the too localized matching, and -- to a lesser degree --
the use of only 3 QNMs). Note, however that (as exhibited
in the inset) this ``coarser'' EOB-type implementation still succeeds in following
the phase of the exact signal to a rather impressive $\pm3.5\%$ of a cycle.

\section{Conclusions}
\label{sec:conc}
We have presented a first attack on the problem of computing analytic template
waveforms, describing the complete binary black hole coalescence phenomenon 
(from early inspiral to late ring-down) that are both {\it effectual} and {\it faithful} 
(in the terminology of Ref.~\cite{Damour:1997ub}).
 We have restricted our analysis to the small-mass-ratio 
 limit without spin, a problem that is amenable of a semi-numerical 
investigation based on a synergic use of black hole perturbation theory and 
PN-based EOB results~\cite{Nagar:2006xv}.

Our analytic approach is based on the Effective One Body (EOB) 
method~\cite{Buonanno:1998gg,Buonanno:2000ef, Damour:2000we,Damour:2001tu,Damour:2006tr}.
The current implementations \cite{Buonanno:2000ef,Buonanno:2005xu,Damour:2006tr,Buonanno:2006ui,Pan:2007nw}
of this method have been recently shown \cite{Pan:2007nw} to define {\it effectual} templates for
the coalescence of comparable-mass (non-spinning) systems in the full mass range
of interest for ground-based GW detectors.
Our aim is to {\it improve} the implementation of the EOB method so as to construct
{\it high-accuracy, faithful} templates. Our strategy towards this aim is to 
understand, element by element, the physics behind each feature of the waveform,
by comparing generalized, multi-parameter EOB-based waveforms to ``exact'' waveforms
obtained by numerical relativity approaches. In this first paper, we have
tapped the information contained in small-mass-ratio waveforms, which can rather
easily be computed very fast and with high-accuracy.

Our main achievements are: (i) a better description of radiation reaction
during the plunge (by taking into account non-quasi-circular effects);
(ii) a refined analytical expression for the $(\ell,m)=(2,2)$ waveform
(which includes a resummation of tail effects, a 3PN-accurate ``non-linear''
amplitude correction, and the inclusion of non-quasi-circular corrections);
 (iii) an improved technique for matching the plunge and
ring-down waveforms (which is based on an improved understanding of the physical
elements entering this matching: origin of the excitation of quasi-normal modes, 
sign of their frequencies, use of an extended matching interval). The implementation
of these improvements has led us to
 the explicit computation of EOB-type analytic
waveforms that accurately reproduce (both in its phase and its amplitude)
the exact waveform during the whole transition. In particular, the 
phase difference stays within $\pm 1.1\%$ of a cycle.
Such a good result in the small mass-ratio case rises the hope that a 
comparably good agreement could also be achieved  in the comparable-mass 
case. This will be the object of future work~\cite{DN07b}.

\acknowledgments
We thank E.~Berti and A. Buonanno for assistance during the conception of 
this work; R.~De~Pietri and L.~Rezzolla for discussions, and
A.~Tartaglia for support. Part of the numerical work has
been carried out by means of the commercial software MATLAB$^{\rm TM}$. 
The MATLAB$^{\rm TM}$ implementation of the {\it complex} $\Gamma$-function
in Eq.~(\ref{eq:tail_term}) is due to Paul Godfrey and can be found at: 
{\tt http://home.att.net/$\sim$ numericana/answer/info/godfrey.htm}.
AN is grateful to IHES and AEI for hospitality during the inception and 
development of this work; he also thanks K.~Kokkotas and ILIAS for 
financial support.

\end{document}